\documentclass[conference]{IEEEtran}
\usepackage{cite}
\usepackage{amsmath,amssymb,amsfonts}
\usepackage{algorithmic}
\usepackage{graphicx}
\usepackage{textcomp}
\usepackage{xcolor}
\usepackage{amsthm}

\newtheorem{lemma}{Lemma}
\newtheorem{theorem}{Theorem}

\usepackage{hyperref}
\usepackage{cite}
\usepackage{amsmath,amssymb,amsfonts}
\usepackage{algorithmic,algorithm}
\newcommand{\rmv}[1]{}
\usepackage[left=1.62cm,right=1.62cm,top=0.73 in,bottom= 1.1in]{geometry}

\begin{document}
\bstctlcite{IEEEexample:BSTcontrol}

\title{Random Alloy Codes and the Fundamental Limits of Coded Distributed Tensors\\
\thanks{Identify applicable funding agency here. If none, delete this.}
}
\author{\IEEEauthorblockN{ Pedro Soto}
\IEEEauthorblockA{
\textit{Virginia Tech} \\
}
}

\maketitle
\begin{abstract}
Tensors are a fundamental operation in distributed computing, \emph{e.g.,} machine learning, that are commonly distributed into multiple parallel tasks for large datasets. Stragglers and other failures can severely impact the overall completion time. Recent works in coded computing provide a novel strategy to mitigate stragglers with coded tasks, with an objective of minimizing the number of tasks needed to recover the overall result, known as the recovery threshold. However, we demonstrate that this strict combinatorial definition does not directly optimize the probability of failure.

In this paper, we focus on the most likely event and measure the optimality of a coding scheme more directly by its probability of decoding. Our probabilistic approach leads us to a practical construction of random codes for matrix multiplication, i.e., locally random alloy codes, which are optimal with respect to the measures. Furthermore, the probabilistic approach allows us to discover a surprising impossibility theorem about both random and deterministic coded distributed tensors.
\end{abstract}
\setlength{\abovedisplayskip}{4.5pt}
\setlength{\belowdisplayskip}{4.5pt}

\section{Introduction}

Machine learning has become a dominant tool in the broader computing community. Its success has been due to the availability of large datasets and more recently the use of hardware optimized to perform multi-linear functions such as GPU's and TPU's,
which are a fundamental building block for a large number of deep neural network architectures.
If the dataset has many data points then the overall computation, or \textit{job}, is distributed as \textit{tasks} amongst \textit{workers}, which model a distributed network of computing devices. This solution creates a new problem; stragglers and other faults can severely impact the performance and overall training time.

The model we consider is the master-worker model of distributed computing where there is a centralized node, \emph{i.e.,} the master, which sends out the computational tasks to the workers and then receives their results. 

This paper offers a novel information-theoretic approach for analyzing the communication complexity of distributed fault-tolerant tensors computation and matrix multiplication in this model. 
Our scheme leads us to construct two families of practical code constructions which we call the \textit{globally random alloy codes} and the \textit{locally random alloy codes}; furthermore, we prove that our constructions give better performance than many benchmarks, \emph{e.g.,} the state-of-the-art algorithms in coded distributed computing.
Our algorithm gives an explicit construction that archives the theoretical bound in \cite{Yu2018StragglerMI} with (arbitrarily) high probability.
We give an implementation that achieves a (probabilistic) bound of $\mathcal{O}(x^{1-\alpha})$ for matrix multiplication of two $x \times x$ block matrices,
 where $\alpha = 0.321334...$ is the current bound for the dual exponent given in \cite{doi:10.1137/1.9781611977912.134}.
We consider schemes that satisfy the weaker but more general condition ``with probability close to 1, results from the first $k$ out of $n$ workers allow the algorithm to terminate correctly''; \emph{i.e.,} this paper considers the \textit{typical fault pattern}.
Our main contributions are: 1) constructing random codes that work for arbitrary fields (cf. \cite{Subramaniam2019RandomKC} which only works for characteristic 0 fields), 2) constructing codes that allow for more than $q$ evaluation points over a finite field, 3) proving that a very large class of outer product codes do not exist over finite fields, and 4) constructing codes that can use less than a factor of 2 many machines in comparison with their un-coded algorithm (cf. \cite{Yu2018StragglerMI} which needs at least 2 times as many machines in comparison to their uncoded counterpart).

\vspace{-8.5pt}
\subsection{Motivating examples: initial demonstration of alloy codes}
Suppose that we wish to multiply matrices
$
  A=  \begin{pmatrix}
    A_{1} & A_{2}  \\
  \end{pmatrix}^T ,
  B=  \begin{pmatrix}
    B_{1} & B_{2}  \\
  \end{pmatrix}  .
  $
If we have $5$ or more workers (\emph{i.e.,} $n\geq 5$), then we can give worker $k$ the coded task
$$ \left(\left(g_A\right)^k_{1}A_1 + \left(g_A\right)^k_{2} A_2 \right) \left(\left(g_B\right)^k_{1}B_1 + \left(g_B\right)^k_{2} B_2 \right)$$ where the $g^k_{i}$
are i.i.d. according to the distribution in Eq.~\ref{eq:non_zero_skew}.
Then the workers will return the values $$(g_C)^k_{i,j}A_iB_j := (g_A)^k_{i}(g_B)^k_{j}A_iB_j.$$ We will see in 
Thm.~\ref{lem:p_adic_lemma} that: the probability of the $(g_C)^k_{i,j}$ being an invertible matrix, where $k$ indexes the rows and the $(i,j)$ indexes the columns, gets arbitrarily close to 1
for a large enough field size $q$; and that this coding scheme works for all fields, \emph{e.g.,} in the case of bytes (\emph{i.e.,} $2^8$) the probability for this example is 0.99 (\emph{i.e.,} it gets close to 1 fast).  In particular, any 4 workers allow us to solve the linear system given by $g_C$. We will see that the coefficients $ (g_A)^k_{i}, (g_B)^k_{j}, (g_C)^k_{i,j}$ form what we term a {\it globally random alloy code} and satisfy the relationship:
$$
 \resizebox{\hsize}{!}{$
g_C = \begin{pmatrix}
(g_A)^1_{1 } (g_B)^1_{1}& (g_A)^1_{1}  (g_B)^1_{2} & (g_A)^1_{2 } (g_B)^1_{1}& (g_A)^1_{2}  (g_B)^1_{2}\\
(g_A)^2_{1 } (g_B)^2_{1}& (g_A)^2_{1}  (g_B)^2_{2} & (g_A)^2_{2 } (g_B)^2_{1}& (g_A)^2_{2}  (g_B)^2_{2}\\
\vdots & \vdots & \vdots & \vdots \\
(g_A)^5_{1 } (g_B)^5_{1}& (g_A)^5_{1}  (g_B)^5_{2} & (g_A)^5_{2 } (g_B)^5_{1}& (g_A)^5_{2}  (g_B)^5_{2}\\
\end{pmatrix}.$}
$$

To illustrate the simplest form of \textit{locally random  alloy code}, suppose that instead want to multiply the following:
\begin{equation*}
 \resizebox{0.9\hsize}{!}{$
  A=  \begin{pmatrix}
    A_{1,1} & A_{2,1} &   A_{3,1} & A_{4,1} \\
    A_{1,2} & A_{2,2} &   A_{3,2} & A_{4,2} \\
  \end{pmatrix}^T ,   
  B=  \begin{pmatrix}
    B_{1,1} & B_{1,2}  & B_{1,3} & B_{1,4}  \\
    B_{2,1} & B_{2,2}  & B_{2,3} & B_{2,4}  \\
  \end{pmatrix}  .
 $}
\end{equation*}
  If we let
  $$
\begin{array}{ll}
       A^1 =  \begin{pmatrix}
      A_{1,1} & A_{2,1}  \\
    \end{pmatrix}^T ,
    & 
    A^2=  \begin{pmatrix}
      A_{1,2} & A_{2,2} \\
    \end{pmatrix} ^T,
    \\
      A^3=  \begin{pmatrix}
      A_{3,1} & A_{4,1} \\
     \end{pmatrix} ^T,
     &
      A^4=  \begin{pmatrix}
      A_{2,3} & A_{2,4} \\
    \end{pmatrix}^T,
   \\
      B^1=  \begin{pmatrix}
      B_{1,1} & B_{1,2}  \\
    \end{pmatrix} ,
   &
    B^2=  \begin{pmatrix}
      B_{1,3} & B_{1,4} \\
    \end{pmatrix}
  ,
  \\
      B^3=  \begin{pmatrix}
      B_{2,1} & B_{2,2} \\
     \end{pmatrix} ,
     &
      B^4=  \begin{pmatrix}
      B_{2,3} & B_{2,4} \\
    \end{pmatrix},
  \end{array}$$
then Strassen's algorithm gives   $AB = T_1+\cdots+T_7$, where
$$
\begin{array}{ll}
T _{1}=({A}^1+{A} ^4)({B}^1+{B} ^4)
, &
T _{2}=({A} ^3+{A} ^4){B}^1, \\
T _{3}={A}^1({B} ^2-{B} ^4), &
T _{4}={A} ^4({B} ^3-{B}^1),
\\
T _{5}=({A}^1+{A} ^2){B} ^4, &
T _{6}=({A} ^3-{A}^1)({B}^1+{B} ^2), \\
T _{7}=({A} ^2-{A} ^4)({B} ^3+{B} ^4).
\end{array}
$$
However, $A'={A}^1+{A} ^4$ and $
B'={B}^1+{B} ^4$ can  be further partitioned and we can apply the code from the previous example to get coded tasks $$ \tilde{A'}\cdot \tilde{B'} = \left(\left(g_A^1\right)^k_{1}A_1' + \left(g_A^1\right)^k_{2} A_2' \right)\cdot\left(\left(g_B^1\right)^k_{1}B_1' + \left(g_B^1\right)^k_{2} a_2' \right).$$
 Also,  $A''={A} ^3+{A} ^4$ and $B''={B}^1$ can be further encoded:$$ \tilde{A''}\cdot \tilde{B''}  = \left(\left(g_A^2\right)^k_{1}A_1'' + \left(g_A^2\right)^k_{2} A_2'' \right)\cdot\left(\left(g_B^2\right)^k_{1}B_1'' + \left(g_B^2\right)^k_{2} a_2'' \right)$$ and continuing for each $T_i$ gives 7 codes $$(g_A^t)^k_{i}, (g_B^t)^k_{j}, (g_C^t)^k_{i,j}, \ t\in [7]$$ which allows us to solve for the $T_i$ from about 28 workers (or more) with high probability; not all size 28 subsets will solve for the $T_i$, but nonetheless, the typical number of workers needed still ends up being less than Entangled Polynomial (EP) codes \cite{Yu2018StragglerMI} which needs a minimum of 33 workers in this case.
 
  Furthermore, the master can drop communication with workers returning useless data so that the master always can receive a maximum of 28 results from the workers. For instance, suppose that the total number of workers is 35. We will see that the recovery threshold for the locally random alloy code is 34, since there exists a fault pattern that returns say 5 workers from groups $T_1 $ through $T_6$ and only 3 from the last group $T_7$; however, fault patterns of this type are highly improbable. Conversely, there are fault patterns where $28,29,30,31,32,34$ and $35$ workers can compute the product; 
  however, for the EP codes \cite{Yu2018StragglerMI}, only the patterns with $33,34,35$ workers returning will work. 

In this paper, we describe codes inspired by these examples. Before doing so, we set the notation to be used throughout.

\subsection{Background}

If $A$ is a $x \times z$ matrix with entries in $\mathbb{F}^{P \times R}$ (equivalently, $A$ is a $xP \times zR$ matrix that has been partitioned into $x \times z$ parts) and $B$ is a $z \times y$ matrix with entries in $\mathbb{F}^{R \times Q}$, 
then, if $T$ is the matrix multiplication operator takes the form $T: (\mathbb{F}^{P \times R})^{x \times z} \times (\mathbb{F}^{R \times Q})^{z \times y} \rightarrow (\mathbb{F}^{P \times Q})^{x \times y}$, then we define the tensor rank as the smallest number $r$ of 
$x \times z$-dimensional linear functionals $E^i_{0}: (\mathbb{F}^{P \times R})^{x \times z} \rightarrow \mathbb{F}^{P \times R}$,  $z \times y$-dimensional linear functionals $E^i_{1}:  (\mathbb{F}^{R \times Q})^{z \times y} \rightarrow \mathbb{F}^{R \times Q}$, and coefficients $D_i \in \mathbb{F}$ such that 
$
    T(A,B) = \sum_{i \in [r]} D_iE^i_0(A)E^i_1(B).
$ 
This modification is a straightforward modification of the usual definition and is implicitly used for recursive fast matrix multiplication algorithms. 
In practice, the parameters $P,Q,R$ are determined by the amount of memory/computing power available at the worker nodes. 

Suppose we are given some family of subsets, $\mathcal{R} \subset [n] $, where $n$ is the number of workers, which we take to be the recoverable subsets. Then we have a family of coefficients 
$D_i^\mathcal{R}$ 
so that the coded matrix multiplication tensor is given by
\begin{equation}\label{eq:coded_tens}
    T_\mathcal{R}(A,B) = \sum_{i \in \mathcal{R}} D_i^\mathcal{R}\tilde E^i_0(A) \tilde E^i_1(B). 
\end{equation}
In particular, the master first encodes (and sends out) $\tilde E^i_0(A), \tilde E^i_1(B)$ for all $i$, then worker $i$ computes $\tilde E^i_0(A) \cdot \tilde E^i_1(B)$, and finally when some recoverable subset $\mathcal{R}$ of workers return, the master decodes by Eq.~\ref{eq:coded_tens}. 
This easily generalizes to higher order products 
$$
    T_\mathcal{R}(A_1,..,A_\ell) = \sum_{i \in \mathcal{R}} D_i^\mathcal{R} \prod_{j \in [\ell]}\tilde E^i_j(A_j) . 
$$

The works \cite{Lee2018a, Yu2017PolynomialCA, 8765375} established the use of coding theory for distributed coded matrix-vector and vector-vector multiplication; shortly after and concurrently, the works \cite{Yu2018StragglerMI, Dutta2018AUC} further improved on these techniques by using alignment to come up with efficient constructions for general matrix-matrix multiplication. 
In particular, the authors of \cite{Yu2018StragglerMI} established the equivalence of the recovery threshold for general partitions to the tensor rank of matrix multiplication (which they call the bilinear complexity). 
The most similar approaches to our work are the codes presented in \cite{Subramaniam2019RandomKC} and the codes in  \cite{Yu2018StragglerMI} as well as the impossibility results in \cite{censorhillel2023nearoptimal}.
The works \cite{8765375, 8006963, 8437549, Dutta2018AUC, pmlr-v80-wang18e, Soto2019DualEP, 8758338, 8849395, hong21b} also worked on improving coded distributed matrix multiplication. 
Further work has been extended to include batch matrix multiplication as well \cite{pmlr-v89-yu19b, 9149322, 9174239, 9750133}.
The work \cite{DOliveira2019GASPCF} considered the extra constraint of private/secure multiplication and the work \cite{Tang2019ErasureCF} considered a trade-off between the size of the entries of the matrix and the recovery threshold, and parallel integer multiplication \cite{schartz24}.
The works \cite{pmlr-v70-tandon17a, pmlr-v162-soto22a} considered coding for distributed gradient descent in machine learning applications.

The channel model used for the matrix multiplication problem is defined as a probability distribution $p(Y = c | X = a)$ where
the random variables $X$ and  $Y$ take values in the sets
 $   \mathcal{X} = \{(a_1,a_2)\ |\ a_1 \in  \mathbb{F}_q^{PS}
,\ a_2 \in  \mathbb{F}_q^{SQ}
\} \cong \mathbb{F}_q^{PS+SQ}
$, $
      \mathcal{Y} = \{c = a_1\cdot a_2\ |\ a_1 \in  \mathbb{F}_q^{PS}
,\ a_2 \in  \mathbb{F}_q^{SQ}
\} \cup \{E\} \subset \mathbb{F}_q^{PQ} \cup \{E\},$
$E$ is a placeholder symbol for an \textit{erasure} (\emph{i.e.,} when a node in the network has a fault),
 $c,a_1,a_2$ are all matrices, and $\cdot $ is matrix multiplication. We call
$
    \mathcal{M} = \{(M_1,M_2)\ |\ M_1 \in  \mathbb{F}_q^{(xP)(zS)}
,\ M_2 \in  \mathbb{F}_q^{(zS)(yQ)}
\} \cong \mathbb{F}_q^{z(xPS+ySQ)}
$ and
$\mathcal{M}' = \{M_1 \cdot M_2\ |\ M_1 \in  \mathbb{F}_q^{(xP)(zS)}
,\ M_2 \in  \mathbb{F}_q^{(zS)(yQ)}
\} \subset \mathbb{F}_q^{(xP)(yQ)}
$
  the input message set and output message set respectively.\footnote{$\mathcal{M}$ is the set of matrices that we wish to multiply and $\mathcal{M}'$ is their product; \emph{i.e.,} the values of $\mathcal{M}$ are $(A,B)$ where $A$ is an $x \times z$ block matrix with blocks of size $P \times R$ and $B$ is an $z \times y$ block matrix with blocks of size $R \times Q$. The product has dimensions $x \times y$ with blocks of size $P \times Q$.}
A code is a pair of functions $(\mathcal{E} ,\mathcal{D} )$ where $\mathcal{E} :\mathcal{M} \longrightarrow \mathcal{X}^n $ is the encoder and $\mathcal{D} : \mathcal{Y}^n \longrightarrow \mathcal{M}'$ is the decoder.
The meaning of $n$ is that we break down a large task $M_1M_2$ into a sequence of $n$ multiplications $a^1_1a^1_2,a^2_1a^2_2,...,a^n_1a^n_2$ which corresponds to using the channel $n$ times.
We assume that each worker's failure probability, $p_f$, is i.i.d; which is a weak assumption in a distributed environment with simple 1-round master-to-worker communication.
For simplicity, we define \textit{Random codes} as probability distributions over all possible codes $\mathcal{C} = (\mathcal{E},\mathcal{D})$. 
The deterministic case is a special case of the non-deterministic case.
\textit{Tensor codes} for matrix multiplication are codes of the form $\mathcal{E} = g_A,g_B$ and $\mathcal{D}$ which solves the equation $g_C(A\cdot B)_\mathrm{flat} = M$ where $g_A*g_B = g_C$, $A,B$ are block matrices, and $_\mathrm{flat}$ converts a matrix to a vector; \emph{i.e.,}
\begin{equation}\label{eq:mat_mult_code}
    (g_C)^k_{i,j}= (g_A)^k_{i}(g_B)^k_{j},
    \end{equation}
    where $k$ corresponds to the $k^\mathrm{th}$ worker.
    % 2024 delete start
    For higher order tensors the code is defined as
    \begin{equation}\label{eq:tensor_code}
    (g_C)^k_{i_1,\dots , i_l}= (g_{A_1})^k_{i_1}...(g_{A_l})^k_{i_l}.
    \end{equation}
    \textit{It is important to note that the operation $*$ is not matrix multiplication but instead defined by Eq.~\ref{eq:mat_mult_code}}.

Given $\mathcal{C} = (\mathcal{E},\mathcal{D} )$ and the channel above, we define our \textit{probability of error} as
$
    p^{\mathcal{C}}_e (M_1,M_2) : = p(\mathcal{D}(Y^n) \neq M_1M_2 \ | \ X^n  = \mathcal{E}(M_1,M_2) ).
$
A code $\mathcal{C}$ with $M=|\mathcal{M}|$ input messages, codewords of length $n$, and probability of error $\epsilon $ is said to be a $(M,n,\epsilon )$-code or $(M,n)$-code for short.
We give the definition of the \textit{(computational) rate} of a $(M,n)$-code as
$
     R = \frac{\log(|\mathcal{M'}| - |\mathcal{E}|)}{ n}  =  \frac{\log(|\mathcal{M'\setminus \mathcal{E}}|)}{ n}
$
where $\mathcal{E}$ is the set of output sequences with an erasure.

The \textit{typical recovery threshold}, $\mathcal{R} (x,y,z,p_f,\epsilon , \mathcal{C})$, is the number of workers needed for the code $\mathcal{C
}$ to have less than $\epsilon $ probability of error for a partition of type $(x,y,z)$ if each worker has a probability of fault equal to $p_f$.
The typical recovery threshold and the rate satisfy the following relationship
$
     \mathcal{R} =   \frac{\log(|\mathcal{M}'\setminus \mathcal{E}|)}{ R}  = \log(|\mathcal{Y}|-1) \frac{xy}{ R}.
$

\section{Code Constructions and Analysis}
Before giving the main construction we give the key lemma of this paper which computes the rank of a matrix whose rows are given by the flattening of rank 1 tensors.
\begin{theorem}\label{thm:main_thm}
    Suppose that $(g)^{1},...,(g)^{d}$ are i.i.d. uniformly random rank 1 tensors where $d = d_1\cdot ...\cdot d_\ell$, and $(g)^{k} \in \mathbb{F}^{d_1} \otimes ... \otimes \mathbb{F}^{d_\ell}$. Let $G$ be the matrix formed by flattening the $(g)^{i} $ and so that the $i^\mathrm{th}$ row of $G$ is $\mathrm{flat}((g)^{k})$. In particular, in the case where $\ell = 2$, we have that 
    $$
   (G)_{k,id_2+j} = (g)^k_{i,j}
    $$
    if we use standard matrix notation. 
    If $d_1+...+d_\ell \geq d_1\cdot ... \cdot d_\ell$,
    then we have that the probability that $G$ is full rank is greater than or equal to 
    \begin{equation}
         \prod_{i \in [d]}  \left[\left(\prod_{j \in [\ell]}1-q^{ -d_j }\right)  -q^{i-d_1-...-d_\ell} \right].
    \end{equation}
\end{theorem}

\begin{proof}
For simplicity let us abuse notation and identify the tensor $g$ with its flattening $G$.
    Let $\mathcal{N}(k,d)$ be the number of $k \times d$ full rank matrices whose rows are (a flattening of) rank 1 tensors (where $k \leq d$).  
   We proceed to prove that 
   \begin{equation*}
       \mathcal{N}(k,d) \geq \prod _{i \in [k]}\left[\left(\prod_{j \in [\ell]}q^{ d_j }-1\right) - q^i \right]
   \end{equation*}
   by induction. 
For the base case, a $1 \times d$ matrix $g$ is full rank iff its first and only row $(g)^1$ is a non zero vector. 
There are a total of 
$
\mathcal{N}(1,d) = \prod_{j \in [\ell]}(q^{ d_j }-1)
$
non-zero rank 1 vectors since $(g)^1_{i_1,...,i_\ell} = (g_{1})^1_{i_1} \cdot ... \cdot (g_{\ell})^{1}_{i_\ell} $ for some $g_{i} \in \mathbb{F}^{d_i}$ by definition and there are $q^{d_1}-1$ choices for $(g_{1})^1$, $q^{d_2}-1$ choices for $(g_{2})^1-1$, and so on and we subtract out the 0 vector ($\ell$ times).

Now assume that 
   \begin{equation*}
      \mathcal{N}(k,d) \geq \prod _{i \in [k]}
      \left[\left(\prod_{j \in [\ell]}q^{ d_j }-1\right) - q^i \right]. 
   \end{equation*}
   To make a full rank $k+1$ row matrix we must first choose $k$ many linearly independent rows of flattened rank 1 matrices and then for the choice of the $(k+1)^\mathrm{th}$ row we must subtract out all possible linear combinations of the previous $k$ rows; this is bounded by $q^{k+1}$.
   There are $ 
          (\prod_{j \in [\ell]}q^{ d_j }-1) - q^{k+1}
  $
  many such linear combinations and thus an induction gives us that
   \begin{multline*}
       \mathcal{N}(k+1,d) \geq  
       \mathcal{N}(k,d) (\prod_{j \in [\ell]}q^{ d_j }-1) - q^{k+1}) 
       \\ \geq \prod _{i \in [k+1]}
      \left[\left(\prod_{j \in [\ell]}q^{ d_j }-1\right) - q^i \right]. 
   \end{multline*}
   Finally the proof is completed by dividing the number of full rank matrices by the number of total matrices whose rows are flattened rank 1 tensors.
   Indeed, we have that the total number is equal to 
   \begin{equation*}
       \mathcal{T}(k,d) =( 
       q^{d_1+...+d_\ell})^k = \prod_{i \in [k]} q^{d_1+...+d_\ell}
   \end{equation*}
  since there are $q^{d_1+...+d_\ell}$ many choices for a row and each of the rows can be chosen independently and we have $k$ many rows where we repeat the same sequences of choices and thus taking the ratio of the two we get the probability is equal to 
   \begin{multline*}
       \frac{\mathcal{N}(k,d)}{\mathcal{T}(k,d)} = \prod_{i \in [k]}\frac{\left(\prod_{j \in [\ell]}q^{ d_j }-1\right) - q^i}{q^{d_1+...+d_\ell}} 
       \\ =   \prod_{i \in [k]}  \left[\left(\prod_{j \in [\ell]}1-q^{ -d_j }\right)  -q^{i-d_1-...-d_\ell} \right] , 
   \end{multline*}
   as was needed to be shown.
\end{proof}

\subsection{Globally Random Alloy Codes for Order 2 Outer Products}
Consider the \textbf{uniformly random 2-product distribution} 
\begin{equation}\label{eq:non_zero_skew}
    p^*(g^i_j = z) = \begin{cases}
  1 - \sqrt{\frac{q-1}{q}} & \text{for } z = 0  \\
    \frac{1}{\sqrt{q(q-1)}} & \text{for } z \neq 0. \\
  \end{cases}
\end{equation}
Then we can define our random codes as
\begin{equation}\label{eq:mat_code}
\tilde{A}_k = \sum_{i \in [x]} (g_A)^k_iA_i, \ \ \ \tilde{B}_k = \sum_{i \in [y]} (g_B)^k_iB_i,
\end{equation}
for matrix multiplication.
The intuition behind the code construction  is that we wish to define probability distributions for the coefficients of $g_A$ and $g_B$ in such a fashion that the resulting product code $g_C$ is i.i.d. uniformly randomly distributed.
If one were to naively encode the coefficients of $g_A$ and $g_B$ using the uniformly random distribution, then the resulting code $g_C$ would be skewed towards 0.
This is because 0 is not invertible and thus multiplication by zero does on form a permutation on $\mathbb{F}_q$.
By slightly skewing $g_A$ and $g_B$ towards non-zero coefficients, we get that the resulting coefficients of the code $g_C$ is i.i.d. by the uniform random distribution.
Since the decodability of the code only depends on the invertibility of $g_C$, the only important distribution is the resulting product distribution; Thm.~\ref{lem:p_adic_lemma} gives us that this is exactly the case.
\begin{theorem}[Existence of Random Tensor Codes for Outer Products of Order 2]\label{lem:p_adic_lemma}
There exists a probability distribution on $g_A,g_B$ such that $g_C = g_A*g_B$ has its coefficients uniformly i.i.d; in particular, the probability of success for scheme over a finite field with $q = p^k$ many elements is $\prod_{i \in [d]}(1-q^{i-d_1-d_2}) $, where $d_1d_2$ is the maximum possible rank of $g_C$.
\end{theorem}

\begin{proof}
Taking the distribution $p^*$ from Eq.~\ref{eq:non_zero_skew} on $g_A , g_B$ gives
$$
   p((g_C)^k_{i,j} = z ) = \sum_{xy=z} p((g_A)^k_{i} = x )p((g_B)^k_{j} = y ),
$$
since  $(g_C)^k_{i,j}= (g_A)^k_{i}(g_B)^k_{j}$ and the $p^*_A,p^*_B$ are independent. Since $p^*_A,p^*_B$ are  identically distributed, we can set
$$
    u = p^*(g^i_j = 0) =  1 - \sqrt{\frac{q-1}{q}}
    $$
    $$
 (\forall z \neq 0)  \  v = p^*(g^i_j = z) =   \frac{1}{\sqrt{q(q-1)}},
$$
where we replace $g_A,g_B$ with $g$ above for simplicity (since they are identically distributed).
A simple combinatorial argument gives us that
$$
   p^*((g_C)^k_{i,j}= z) = \begin{cases}
 u^2+2(q-1)uv & \text{for } z = 0  \\
   (q-1)v^2 & \text{for } z \neq 0 \\
 \end{cases}.
$$
Therefore, the probability of a $0$ is equal to
$$
   p^*((g_C)^k_{i,j}= 0) = u^2+2(q-1)uv =
$$
\resizebox{\hsize}{!}{$
   \left(1 - \sqrt{\frac{q-1}{q}} \right)^2+2(q-1)\left(1 - \sqrt{\frac{q-1}{q}} \right)\left( \frac{1}{\sqrt{q(q-1)}} \right) =  \frac{1}{q}.
   $}
Similarly, we have that
$$
p^*((g_C)^k_{i,j} = z) = (q-1)v^2 =(q-1)\left(\frac{1}{\sqrt{q(q-1)}}  \right)^2 = \frac{1}{q},
$$
for a non-zero $z$.
Because $p^*((g_C)^k_{i,j}= z) $ is a function of the values $p^*((g_A)^k_{i}= 0),...,p^*((g_A)^k_{i}= q-1)$ and $p^*((g_B)^k_{j}= 0),...,p^*((g_B)^k_{j}= q-1)$ we have that $k \neq k'$ implies that $p^*((g_C)^k_{i,j}= z) $ and $ p^*((g_C)^{k'}_{i,j}= z)$ are independent.
Thus, any two rows of $g_C$ are independent.
The proof is completed by applying Theorem~\ref{thm:main_thm}, but in order to apply this theorem we must first show that our code does indeed generate random rank 1 tensors uniformly. 
If either $i \neq i'$ or $j\neq j'$, then $p^*((g_C)^k_{i,j}= z) $ and $ p^*((g_C)^{k}_{i',j'}= z)$ are independent.
Thus if we fix an $i$ (or respectively fix a $j$) we have that $p^*((g_C)^k_{i,j}= z)$ is a uniformly randomly generated over $\mathbb{F}^{d_1}$ (respectively $\mathbb{F}^{d_2}$). 
Since this is true as we fix any $i$ or $j$ we have that for a fixed $k$ every possible rank 1 tensor is generated with equal probability and Theorem~\ref{thm:main_thm} completes the proof.
\end{proof}

\subsection{General Globally Random Alloy Codes of Order $\ell$}
If we let the \textit{uniformly random $l$-product tensor distribution}, be defined as
\begin{equation}\label{eq:non_zero_skew_tensor}
    p^*(g^i_j = z) = \begin{cases}
  1 - \sqrt[l]{\frac{q-1}{q}} & \text{for } z = 0  \\
    \frac{1}{\sqrt[l]{q(q-1)^{l-1}}} & \text{for } z \neq 0 \\
  \end{cases},
\end{equation}
then we can define our random codes as
\begin{equation}\label{eq:tens_code_alg}
(\tilde A_1,..., \tilde A_l )  = \Bigg(\sum_{i \in [a_1]} (g_{A_1})^k_iA_{1,i}, ...,\sum_{i \in [x_l]} (g_{A_l})^k_iA_{l,i}\Bigg).
\end{equation}

\begin{theorem}[Existence of Random Tensor Codes for Outer Products of Order $\ell$]\label{lem:p_adic_lemma_gen}
There exists a probability distribution on $g_{A_1},...,g_{A_\ell}$ such that $g_C = g_{A_1}* ... *g_{A_\ell}$ has its coefficients uniformly i.i.d; in particular, the probability of success for scheme over a finite field with $q = p^k$ many elements is $\prod_{i \in [d]}(1-q^{i-d_1-...-d_\ell}) $, where $d_1\cdot ... \cdot d_\ell$ is the maximum possible rank of $g_C$.
\end{theorem}

\begin{proof}
Taking the distribution $p^*$ from Eq.~\ref{eq:non_zero_skew_tensor} on $g_{A_j} $ gives
$$
   p((g_C)^k_{i_1,...,i_\ell} = z ) = \sum_{x_1\cdot ... \cdot x_\ell=z} \prod_{j \in [\ell]}p((g_{A_j})^k_{i_j} = x_{i_j} ),
$$
since  $(g_C)^k_{i_1,...,i_\ell}= (g_{A_1})^k_{i}(g_C)^k_{j}$ and the $p^*_A,p^*_B$ are independent. Since $p^*_A,p^*_B$ are  identically distributed, we can set
$$
    u = p^*(g^i_j = 0) =  1 - \sqrt[l]{\frac{q-1}{q}} 
    $$
    $$
 (\forall z \neq 0)  \  v = p^*(g^i_j = z) =   \frac{1}{\sqrt[l]{q(q-1)^{l-1}}},
$$
where we replace the $g_{A_i}$ with a $g$ above for simplicity (since they are identically distributed).
A simple combinatorial argument gives us that
\begin{multline}
   p^*((g_C)^k_{i_1,...,i_\ell}= z) 
   \\ = \begin{cases}
 \sum_{i \in [\ell]} \binom{\ell}{i}u^{\ell - i}(q-1)^iv^i & \text{for } z = 0  \\
   (q-1)^{\ell-1}v^\ell & \text{for } z \neq 0 \\
 \end{cases}.
\end{multline}
Therefore, the probability of a $0$ is equal to
$$
   p^*((g_C)^k_{i_1,...,i_\ell}= 0) 
$$
$$
=   (u + (q-1) v)^n-((q-1) v)^n  = 1 - \frac{q-1}{q} = \frac{1}{q}
$$
Similarly, we have that
\begin{multline}
 p^*((g_C)^k_{i_1,...,i_\ell}= z) = (q-1)^{\ell-1}v^\ell \\ 
 =(q-1)^{\ell-1}\left(\frac{1}{\sqrt[\ell]{q(q-1)^{\ell-1}}}  \right)^{\ell} = \frac{1}{q},
\end{multline}
for a non-zero $z$.
Because $p^*((g_C)^k_{i_1,...,i_\ell}= z) $ is a function of the values $p^*((g_{A_1})^k_{i_1}= 0),...,p^*((g_{A_1})^k_{i_1}= q-1), ...,p^*((g_{A_\ell})^k_{i_\ell}= 0),...,p^*((g_{A_\ell})^k_{i_\ell}= q-1)$ we have that $k \neq k'$ implies that $p^*((g_C)^k_{i_1,...,i_\ell}=  z) $ and $ p^*((g_C)^{k'}_{i_1,...,i_\ell}= z)$ are independent.
Thus, any two rows of $g_C$ are independent.
The proof is completed identically to the proof of Theorem~\ref{lem:p_adic_lemma}.
\end{proof}

\begin{lemma}\label{lem:extend_code}
    The codes given by Thm.~\ref{lem:p_adic_lemma} exist for all fields. 
\end{lemma}

\begin{proof}
    Suppose $\mathrm{char}(\mathbb{F}) = p$ for some prime $p$ and  $|\mathbb{F}| > p^{k}$. Then $\mathbb F$ contains a subfield isomorphic to $\mathbb{F}_q$, for all $q = p^{k'}$ and $k'|k$, which is the setting described in Theorem \ref{lem:p_adic_lemma}. 
    Thus, it remains to prove the result for $\mathrm{char}(\mathbb{F}) = 0$. In this case, $\mathbb F$ contains a copy of $\mathbb Z$. 
    Because the determinant function and the function $\mathrm{mod}_p : \mathbb{Z} \rightarrow \mathbb{F}_p$ defined by $ n \mapsto (n \mod p )$ are homomorphisms, 
   a full rank matrix $G$ over $\mathbb{F}_p$ may be considered as a full rank matrix $g_\mathbb{Z}$ with entries in $\mathbb{Z}$, reducing to the previous case. 
\end{proof}

\subsection{An Impossibility Theorem}

\begin{theorem}\label{thm:imp}
 There do not exist linear outer product codes (of neither the deterministic nor random type) if $\left(\prod_{j \in [\ell]}1-q^{ -d_j }\right)  \leq q^{1-d_1-...-d_\ell}\cdot  d_1\cdot...\cdot d_\ell $.
\end{theorem}

\begin{proof}
  A modification of the proof of Theorem~\ref{thm:main_thm} gives us that the total number of full rank $(k+1 )\times d$ matrices that can be created by flattening outer products satisfies the recursion
   \begin{equation*}
       \mathcal{N}(k+1,d) \leq  
       \mathcal{N}(k,d) \left(\left(\prod_{j \in [\ell]}q^{ d_j }-1\right)- q(k+1)\right),   
   \end{equation*}
      which counts the number of $k+1$ linearly independent sets of flattened rank 1 matrices by taking a set of $k$ many such matrices and subtracting out all the possible linearly dependent choices.
However, plugging in $d$ gives us
$
       \mathcal{N}(d,d)/ \mathcal{T}(d,d)\leq  
      0,   
$
and thus there cannot exist $d$ many linearly independent rank 1 tensors if  $\left(\prod_{j \in [\ell]}1-q^{ -d_j }\right)  < q^{1-d_1-...-d_\ell}\cdot  d_1\cdot...\cdot d_\ell $. 
\end{proof}

\subsection{Locally Random Alloy Codes}\label{sec:la_code}

Given a tensor $T$ with a tensor decomposition  $T_1,...,T_r$, and general block matrices $A_0,A_1$, where $T=T_1+...+T_r$, $T_i(A)= D_i(\mathcal{T}_i(E^i_0(A_0) , E^i_1(A_1) )) $, and the workers perform outer products $\mathcal{T}_t$ for a non-coded algorithm, we show how to convert it into a coded algorithm. 
 The master node generates $r$ many codes $g^t:=(g_{A_j}^t)^k_i$ for $t \in[r]$ according to Eq.~\ref{eq:non_zero_skew}.
 The workers are indexed by $(t,k)$ where $t \in [r]$, $k \in [n/r]$.  
 The master and sends worker $(t,k)$ the data $\widetilde A_{k,0}^t ,\widetilde A_{k,1}^t $, where $\widetilde A_{k,j}^t =  \sum_{i \in [\text{dim}(A_j)]} (g_{A_j}^t)^k_iE^t_j(A_{j,i})$.
  Each worker $(t,k)$ computes $\mathcal{T}_t(\widetilde A_{k,0}^t,  \widetilde A_{k,1}^t)$.
  The master then solves for $\mathcal{T}_i(E^i_0(A_0) , E^i_1(A_1) )$ for each $t \in  [r]$ by inverting $g_C^t$ (if possible) and returns $(D_1T_1+...+D_rT_r )(A_0,A_1)$.

\begin{theorem}[Theorem for General 3-D Distributed Matrix Multiplication]\label{lem:3d}
If $R = \frac{p_f}{1-p_f} x^{3 - \alpha } - \epsilon =  \mathcal{O}(x^{2.687...})$,
 where $ \alpha > 0.321334 $ is the bound for dual exponent of matrix multiplication given in \cite{doi:10.1137/1.9781611977912.134}, then local random  alloy codes achieves the rate $R $ for large enough $x= y = z$ and $p_f \geq 1/2$. For the simpler case where $p_f < 1/2$ one can replace $\frac{p_f}{1-p_f}$ with $\frac{1-p_f}{p_f}<2$. 
\end{theorem}

\begin{proof}
Let $T= T_1+...+ T_r$ be an optimal tensor rank decomposition for $(\frac{x}{\mu},y,z)=(\frac{z}{\mu},z,z)$ then we can apply the code in Eq.~\ref{eq:mat_code} to the $T_i$ to get $r$ optimal $(\mu,1,1)=(z^{1-\alpha},1,1)$ codes.
Then this creates a block code for $T$ via standard arguments by the method of strong types, \emph{i.e.,} Hoeffding's inequality, as found in \cite{Csiszr1997InformationT}.

To simplify the proof we prove it for a probability of failure $p= p_f \geq \frac{1}{2}$. If we give each worker group a factor of $\lambda$ extra workers and we want to know what is that probability that a percentage of $(p +\Delta )\%$ of them will fail, then we have that
$$
P(\text{$(p +\Delta )\%$ of workers will fail}) \leq \exp\left(-\frac{\Delta ^2\chi}{2p\left(1-p\right)}\right)
$$
by the Chernoff-Hoeffding inequality.
We need
$
p+\Delta  = \frac{\lambda z^{1-\alpha}}{(1+\lambda ) z^{1-\alpha}} = \frac{\lambda}{1+\lambda }
$
so that
$
\Delta =  \frac{\lambda}{1+\lambda } - p.
$
Therefore if we want the probability of the $z^2$ larger tasks $T_1,...,T_{z^2}$ tasks (see \cite{doi:10.1137/1.9781611975031.67} for a proof that $z^2$ rank-one tensors are needed for $(\frac{z}{\mu},z,z)$ size rectangular multiplication) to fail with less than $1- \epsilon $ probability the following bound
$$
\left(1-\exp\left(-\frac{\left(\frac{\lambda}{1+\lambda } - p\right)^2\chi}{2p\left(1-p\right)}\right)\right)^{z^2}  \geq 1- \epsilon,
$$
is sufficient.
Algebraic manipulation and the facts that 
$$
\exp\left(-\frac{\left(\frac{\lambda}{1+\lambda } - p\right)^2\chi }{2p\left(1-p\right)}\right) < 1
$$
and  
$c <1 \implies 1-c \geq \log (1-c)$ 
together imply that
$$
-\exp\left(-\frac{\left(\frac{\lambda}{1+\lambda } - p\right)^2\chi}{2p\left(1-p\right)}\right)
 \geq
 \frac{\ln\left( 1- \epsilon \right)}{ z^2 } - 1.
$$
which in turn gives 
$
-\frac{\left(\frac{\lambda}{1+\lambda } - p\right)^2\chi}{2p\left(1-p\right)}
 \leq
 \ln\left(1- \frac{\ln\left( 1- \epsilon \right)}{ z^2 } \right)
$
and 
$$
\frac{\lambda}{1+\lambda }
 \geq
 p+ \sqrt{-\chi^{-1}2p\left(1-p\right)\ln\left(1- \frac{\ln\left( 1- \epsilon \right)}{ z^2 } \right)} .
$$
Fixing $p, \epsilon$ and letting $x,y,z,\chi \to \infty$ gives us that for any $\epsilon ' \to  0  $, the inequality
 $
 \frac{\lambda}{1+\lambda }
  \geq
  p+ \epsilon
 $
 is satisfied. Thus we have that asymptotically we only need that
 $
 \frac{\lambda}{1+\lambda }
  \geq
  p
 $
 which is equivalent to the desired inequality
 $
\lambda
  \geq
  \frac{p_f}{1-p_f}$.
\end{proof}

\bibliographystyle{IEEEtran}
\bibliography{ref,references, ref2}

\end{document}